\documentclass[twocolumn,superscriptaddress,groupedaddress,reprint,preprintnumbers,amsmath,amssymb,aps,prb,floatfix,nolongbibliography,linenumbers]{revtex4-2}

\usepackage{epsfig}
\usepackage{graphicx}
\usepackage{dcolumn}
\usepackage{bm}
\usepackage{caption}
\usepackage{ragged2e}
\usepackage{ulem}
\usepackage{hyperref}
\usepackage{color,soul}
\usepackage{float}
\hypersetup{colorlinks=true, linkcolor=blue, citecolor=blue,
	filecolor=blue, urlcolor=blue}
\begin{document}
	\nolinenumbers
	
	
	\title {Role of electron correlation and disorder on the electronic structure of layered nickelate (La$_{0.5}$Sr$_{0.5}$)$_2$NiO$_4$}
	
	\author{Sakshi Bansal}
	\affiliation{Department of Physics, Indian Institute of Science
		Education and Research Bhopal, Bhopal Bypass Road, Bhauri, Bhopal
		462066, India}%
	
	\author{R. K. Maurya}
	\affiliation{Department of Physics, Indian Institute of Science
		Education and Research Bhopal, Bhopal Bypass Road, Bhauri, Bhopal
		462066, India}%
	
	\author{Asif Ali}
	\affiliation{Department of Physics, Indian Institute of Science
		Education and Research Bhopal, Bhopal Bypass Road, Bhauri, Bhopal
		462066, India}%
	
	\author{B. H. Reddy}
	\altaffiliation[Present address: ]{Department of Physics, Government
		College (A), Rajahmundry 533105, India}
	\affiliation{Department of Physics, Indian Institute of Science
		Education and Research Bhopal, Bhopal Bypass Road, Bhauri, Bhopal
		462066, India}%
	
	\author{Ravi Shankar Singh}
	\email{rssingh@iiserb.ac.in}
	\affiliation{Department of Physics, Indian Institute of Science
		Education and Research Bhopal, Bhopal Bypass Road, Bhauri, Bhopal
		462066, India}%
	
	
	\begin{abstract}
  We investigate the role of electron correlation and disorder on the electronic structure of layered nickelate (La$_{0.5}$Sr$_{0.5}$)$_2$NiO$_4$ using core level and valence band photoemission spectroscopy in conjunction with density functional theory (DFT) and dynamical mean field theory (DMFT) calculations. Sr 3$d$ and La 4$d$ core level spectra exhibit multiple features associated with photoemission final state effects. An increase of unscreened features in the Sr 3$d$ and La 4$d$ core level spectra with lowering temperature suggests the reduction in density of states (DOS) at the Fermi level, $E_F$. Valence band spectra collected using different photon energies reveal finite intensity at $E_F$ and overall spectra are well captured by DFT+DMFT. Strong renormalization of partially filled $e_g$ bands in DFT+DMFT result indicates strong correlation in this system. Mass enhancement factor, $m^*/m_{\text{DFT}} \sim$ 3, agrees well with values obtained from specific heat measurements. High resolution spectra in the vicinity of $E_F$ show monotonically decreasing spectral intensity with lowering temperature, which evolves to exhibit a Fermi cut-off at low temperatures indicating metallic character in contrast to insulating transport, suggesting Anderson insulating state. $|E-E_F|^{1/2}$ dependence of the spectral DOS and square root temperature dependence of spectral DOS at $E_F$ evidences the role of disorder in the electronic structure of (La$_{0.5}$Sr$_{0.5}$)$_2$NiO$_4$.		
		
	\end{abstract}
	
	
	\maketitle
	
	\section*{INTRODUCTION} 
High $T_c$ superconductivity in strongly correlated layered cuprates \cite{bednorz-physikB-1986sc,*wu-PRL-1987sc,torrance-PRL-1988LSCOsc,*van-PRB-1987LSCOsc,takagi-PRB-1989LSCOsc} have led to exploration of other layered transition metal oxides (TMOs) \cite{imada-review-1998}. Strong electron correlation in 3$d$ TMOs \cite{imada-review-1998} has been regarded as key to the intriguing properties such as superconductivity \cite{bednorz-physikB-1986sc,*wu-PRL-1987sc,torrance-PRL-1988LSCOsc,*van-PRB-1987LSCOsc,takagi-PRB-1989LSCOsc}, giant/colossal magnetoresistance \cite{von-PRL-1993gaint,jin-AIP-1994colossal}, spin/charge/orbital ordering \cite{chen-PRL-1993charge,*cheong-PRB-1994charge,abeykoon-APS-2013charge,anissimova-nature-2014charge,Zaliznyak-PRL2003charge,*Zaliznyak-PRB2015charge,garcia1992-RNiO3-SDW}, \textit{etc}. The recent discovery of superconductivity in hole doped infinite layer nickelates  \cite{li-nature-2019recentsc,hepting-nature-2020recentsc,chen2022-DMFT-RNiO2} has instigated the reinvestigation of physical properties and electronic structure of other layered nickelates. Isostructural to superconducting cuprates (La$_{1-x}$Sr$_x$)$_2$CuO$_4$ \cite{torrance-PRL-1988LSCOsc,*van-PRB-1987LSCOsc,takagi-PRB-1989LSCOsc}, layered nickelates (La$_{1-x}$Sr$_x$)$_2$NiO$_4$ have been studied extensively in search of superconductivity \cite{sreedhar-elsevier-1990LSNO,Ytakeda-elsevier-1990LSNO,RJcava-APS-1991LSNO,liu1991-thermoelectric-LSNOdisorder}.
Both the parent compounds are antiferromagnetic insulators, where La$_2$CuO$_4$ is Mott-Hubbard type as it remains insulating in its paramagnetic state \cite{saylor-APS-1989-LCONeelTn}, while La$_2$NiO$_4$ is Slater-type antiferromagnetic as it becomes metallic above the transition temperature \cite{RJcava-APS-1991LSNO,sreedhar-elsevier-1990LSNO,Ytakeda-elsevier-1990LSNO,schartman-elsevier-1989-LNONeelTn,*yamada-elsevier-1992-LNONeelTn}. A small amount of hole doping (by Sr substitution at La site) leads to disappearance of antiferromagnetic order in both the systems and drives cuprates towards superconducting state \cite{torrance-PRL-1988LSCOsc,van-PRB-1987LSCOsc,takagi-PRB-1989LSCOsc} while nickelates remain insulating \cite {sreedhar-elsevier-1990LSNO,Ytakeda-elsevier-1990LSNO,RJcava-APS-1991LSNO,shinomori-JPS-2002MIT}. Metal to insulator transition (MIT) temperature reduces monotonically with increasing Sr content in (La$_{1-x}$Sr$_x$)$_2$NiO$_4$ and is found to be around 150 K for $x$ = 0.5, while it completely disappears for $x$ $>$ 0.55  \cite{sreedhar-elsevier-1990LSNO,Ytakeda-elsevier-1990LSNO,RJcava-APS-1991LSNO,shinomori-JPS-2002MIT,liu-ACS-1991-composition,granados-elsevier-1993-bandgap,*ishikawa-CSJ-1987-nonstoichiometry,ivanova-springer-2002-transport}. Charge ordering in hole doped cuprates and nickelates have also been studied in relation to disorder introduced by dopant ions \cite{Zaliznyak-PRL2003charge,*Zaliznyak-PRB2015charge}, for example, stripe/checkerboard type charge ordering in (La$_{1-x}$Sr$_x$)$_2$NiO$_4$ for low hole doping \cite{chen-PRL-1993charge,*cheong-PRB-1994charge,abeykoon-APS-2013charge,anissimova-nature-2014charge}. Metallic character observed in optical conductivity suggests that the insulating transport is associated with localization of electrons rather than charge ordering for $x$ = 0.45 and 0.5 \cite{shinomori-JPS-2002MIT}. It is to note here that (La$_{1-x}$Sr$_x$)$_2$NiO$_4$ exhibit different oxygen stoichiometry depending upon the amount of Sr content and synthesis condition, leading to change in the physical properties as seen in the bulk as well as pulsed laser deposited films \cite{sreedhar-elsevier-1990LSNO,Ytakeda-elsevier-1990LSNO,RJcava-APS-1991LSNO,granados-elsevier-1993-bandgap,*ishikawa-CSJ-1987-nonstoichiometry,ivanova-springer-2002-transport,liu-ACS-1991-composition,nitadori-CSJ-1988valence,chouket-elsevier-2016Dielectric,jestadt-PRB-1999Muon,taruti-RSC-2021recent,shinomori-JPS-2002MIT,millburn-elsevier-1999evolution}. When prepared in ambient conditions, (La$_{1-x}$Sr$_x$)$_2$NiO$_4$ with low Sr content ($x$ $\le$ 0.2) can accommodate excess oxygen, while systems with intermediate Sr content are almost stoichiometric and systems with high Sr content ($x$ $>$ 0.55) are found to be slightly oxygen deficient \cite{granados-elsevier-1993-bandgap,*ishikawa-CSJ-1987-nonstoichiometry,ivanova-springer-2002-transport,nitadori-CSJ-1988valence,chouket-elsevier-2016Dielectric,taruti-RSC-2021recent}. Experimental techniques such as thermogravimetric analysis and iodometric titration have confirmed that $x$ = 0.5 system, i.e., (La$_{0.5}$Sr$_{0.5}$)$_2$NiO$_4$ prepared by solid-state reaction route \cite{granados-elsevier-1993-bandgap,*ishikawa-CSJ-1987-nonstoichiometry,chouket-elsevier-2016Dielectric,millburn-elsevier-1999evolution,jestadt-PRB-1999Muon,Ytakeda-elsevier-1990LSNO} is almost stoichiometric.
	
Electronic structure of La$_2$NiO$_4$ has been studied extensively,	from the strong correlation, and/or superconductivity point of view, mostly in the low hole doped regime \cite{szpunar-elsevier-1989-AFMelectronicStr,eisaki-PRB-1992-PES,kuiper-APS-1991unoccupiedDOS,anisimov-APS-1999-electronicStr}. Apart from competitive intra (inter)-site Coulomb interaction, $U$ ($J_H$) and inter-site hopping interaction, disorder also plays a key role in deciding the ground state properties of TMOs, as seen in various systems \cite{reddy-JPCM-palladates,lee-modern-1985disorder,*2021PRBlaSrRhO3films,*kobayashi-PRL-2007AAtheory,*altshuler-elsevier-1979AAtheory,reddy-EPL-2019STIO,singh-PRB-2008YIO,*bansal-JPCM-2021LRO}. The strength of disorder in (La$_{1-x}$Sr$_x$)$_2$NiO$_4$ can be tuned by varying $x$ \cite{liu1991-thermoelectric-LSNOdisorder,liu-ACS-1991-composition,taruti-RSC-2021recent,schilling-IOP-2008-labanio4} where the maximum substitutional disorder is present in the case of (La$_{0.5}$Sr$_{0.5}$)$_2$NiO$_4$. Sommerfeld coefficient of 9.15 mJ/mol/K$^2$ obtained from specific heat measurements \cite{schilling-IOP-2008-labanio4} and room temperature resistivity of $\sim$ 2 - 10 m$\Omega$-cm for (La$_{0.5}$Sr$_{0.5}$)$_2$NiO$_4$ \cite{RJcava-APS-1991LSNO,granados-elsevier-1993-bandgap,ivanova-springer-2002-transport,ishikawa-CSJ-1987-nonstoichiometry,schilling-IOP-2008-labanio4} are very similar to those obtained for LaNiO$_3$ \cite{sreedhar1992La2NiO3}. Thus, both systems, having Ni 3$d^7$ electronic configuration with one electron in $e_g$ band, are expected to have similar order of electron correlation and overall electronic structure \cite{anisimov-APS-1999-electronicStr}. While (La$_{0.5}$Sr$_{0.5}$)$_2$NiO$_4$ exhibits MIT at around 150 K \cite{sreedhar-elsevier-1990LSNO,Ytakeda-elsevier-1990LSNO,RJcava-APS-1991LSNO,shinomori-JPS-2002MIT,liu-ACS-1991-composition,granados-elsevier-1993-bandgap,ivanova-springer-2002-transport,ishikawa-CSJ-1987-nonstoichiometry}, LaNiO$_3$ remains metallic down to low temperatures \cite{sreedhar1992La2NiO3}. Thus, MIT in (La$_{0.5}$Sr$_{0.5}$)$_2$NiO$_4$ may be associated with intrinsic disorder present within the system. Structurally, (La$_{0.5}$Sr$_{0.5}$)$_2$NiO$_4$ consists of NiO$_2$ layer ($ab$ plane) sandwiched between two of the disordered (La/Sr)-O layers from both sides along the $c$-axis leading to local distortion in NiO$_6$ octahedral network \cite{millburn-elsevier-1999evolution}. Thus, (La$_{0.5}$Sr$_{0.5}$)$_2$NiO$_4$ provides an opportunity for understanding the role of electron correlation and disorder on the MIT in this system.

Here, we investigate the electronic structure of layered nickelate (La$_{0.5}$Sr$_{0.5}$)$_2$NiO$_4$ using photoemission spectroscopy and DFT+DMFT calculations. Valence band spectra could be well captured within DFT+DMFT framework, which also suggests strong renormalization of partially filled $e_g$ band ($m^*/m_{DFT} \sim$ 3) due to strong electron correlation. High resolution valence band spectra exhibit a Fermi cut-off at low temperatures, indicating metallic character in contrast to transport measurements suggesting Anderson insulating state. Energy and temperature dependence of spectral DOS following Altshuler-Aronov theory confirms the role of disorder in the electronic structure of strongly correlated disordered nickelate (La$_{0.5}$Sr$_{0.5}$)$_2$NiO$_4$.

\section*{EXPERIMENTAL and CALCULATION DETAILS}

Polycrystalline sample of (La$_{0.5}$Sr$_{0.5}$)$_2$NiO$_4$ was prepared using high purity La$_2$O$_3$, SrCO$_3$ and NiO by solid-state reaction method. Hygroscopic La$_2$O$_3$ was preheated at 900$^o$C for 10 hours before weighing. The well ground mixture was pelletized and sintered at 1300$^o$C for 36 hours with two intermediate grindings. The sample was furnace cooled at the end of each heat treatment. The phase purity and absence of any impurity were confirmed by room temperature powder $x$-ray diffraction (XRD) pattern (Fig. S1 of Supplemental Material (SM) \cite{Supplemental}). The Rietveld refinement of XRD pattern reveals tetragonal crystal structure with space group $I4/mmm$ and lattice parameters $a$ = 3.827 \AA~ and $c$ = 12.452, which are in excellent agreement with the earlier reports \cite{sreedhar-elsevier-1990LSNO,Ytakeda-elsevier-1990LSNO,millburn-elsevier-1999evolution}.
 
Photoemission spectroscopic measurements were performed using Scienta R4000 electron energy analyzer on \textit{in-situ} (base pressure $\sim$ 4 $\times$ 10$^{-11}$ mbar) scraped samples. Total instrumental resolutions were set to 300 meV, 8 meV and 5 meV for measurements with monochromatic Al $K_\alpha$ (1486.6 eV), He {\scriptsize II} (40.8 eV) and He {\scriptsize I} (21.2 eV) radiations (energy), respectively. Ar$^+$ ion sputter cleaned polycrystalline	silver was used to determine $E_F$ and the energy resolutions for different radiations at 30 K. The reproducibility of data and cleanliness of the sample surface were ensured after each trial of scraping by minimizing C 1$s$ feature and the higher binding energy feature in O 1$s$ spectra.
	
Electronic structure calculations were performed using DFT as implemented in WIEN2k \cite{blaha-2001-wien2k,*blaha-2020-wien2k} and fully charge self-consistent DFT+DMFT as implemented in eDMFT code \cite{Haule2010DMFT}. The crystal structure consists of two formula units with a rock-salt arrangement of La and Sr atoms. Experimental lattice parameters were used and atomic positions were relaxed with DFT (forces on each atom $<$~5 meV/\AA). We used the generalized gradient approximation (GGA) for exchange correlation functional \cite{perdew-1996-GGA}. All five Ni 3$d$ orbitals were considered correlated and the local axis was chosen such that $x$ and $y$ axes are aligned towards planar oxygens and the $z$ axis is along apical oxygen. Fully rotational invariant form of Coulomb interaction was used with $U$ and $J_H$ interaction parameters set to 10 eV and 0.7 eV, respectively. Similar values of $U$ and $J_H$ have also been used in case of other nickelates \cite{chen2022-DMFT-RNiO2,assessing2022DMFT}. The temperature was fixed at 116 K ($\beta$ $\approx$ 100 eV$^{-1}$). To subtract double counting (DC), ``exact" DC method was used which provides a better description for correlated materials \cite{haule2015exactDMFT}. Continuous time quantum Monte Carlo impurity solver \cite{haule2007quantumDMFT} was used with 10$^8$ Monte Carlo steps in each run. Maximum entropy method \cite {Haule2010DMFT} for analytical continuation was used to obtain the self energy on the real frequency axis. Mass enhancement factor, $m^*/m_{DFT}$, were obtained from slopes of both Im[$\Sigma$($i\omega$)] (imaginary part of the self energy on imaginary frequency axis) as well as Re[$\Sigma$($\omega$)] (real part of the self energy on real frequency axis) and were consistent \cite{kotliar2DMFT}.
	
	\section*{RESULTS AND DISCUSSION} 
	
	\begin{figure}[tb]
		\centering
		\includegraphics[width=0.48\textwidth]{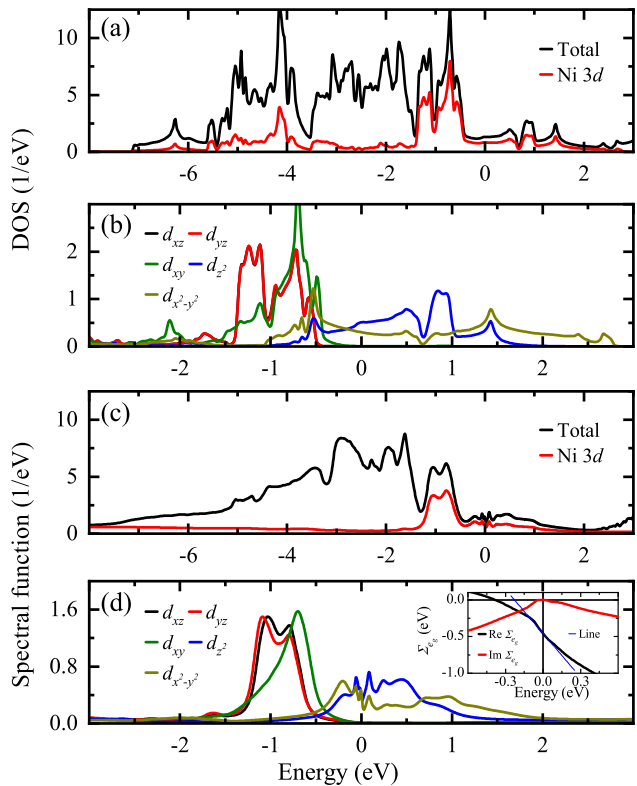}
		\captionsetup{justification=Justified}
		\caption{(color online) (a) Total DOS and Ni 3$d$ partial DOS and (b) orbital resolved partial DOS of Ni 3$d$, for (La$_{0.5}$Sr$_{0.5}$)$_2$NiO$_4$ calculated within DFT. (c) Total and Ni 3$d$ spectral functions and (d) orbital resolved spectral functions of Ni 3$d$, for (La$_{0.5}$Sr$_{0.5}$)$_2$NiO$_4$ calculated within DFT+DMFT. The inset of (d) shows the real and imaginary part of the average self energy of $d_{z^2}$ and $d_{x^2-y^2}$ orbitals. Blue dotted line shows the linear fit to the real part of the self energy in the close vicinity of $E_F$.}\label{fig1}
	\end{figure}

 The results of electronic structure calculation within DFT and DFT+DMFT for (La$_{0.5}$Sr$_{0.5}$)$_2$NiO$_4$ are shown in Fig. \ref{fig1}. Total DOS and Ni 3$d$ partial DOS obtained from DFT are shown in Fig. \ref{fig1}(a). The valence band of this system is mainly formed by the hybridization between Ni 3$d$ and O 2$p$ states as the contributions from La 4$f$, La 5$d$ and Sr 4$d$ states appear mainly above 3 eV energy in the unoccupied region. Total DOS exhibits three sets of features appearing in the energy range, -7 to -3.5 eV corresponding to bonding states, -3.5 to -1.5 eV corresponding to non-bonding O 2$p$ states and -1.5 to 2.5 eV corresponding to anti-bonding states primarily having Ni 3$d$ character. The finite DOS at $E_F$ (DOS($E_F$)) suggests a metallic ground state. Fig. \ref{fig1}(b) shows the orbital resolved Ni 3$d$ band close to $E_F$ where completely filled localized $t_{2g}$ band centers around -1 eV while partially filled delocalized $e_g$ band spans from about -1.5 eV to 2.5 eV and has finite contribution at $E_F$. Slight elongation of bond along apical oxygens in NiO$_6$ octahedra lifts the degeneracy of $t_{2g}$ band to almost degenerate $d_{xz}$/$d_{yz}$ and $d_{xy}$ orbitals, and of $e_g$ band to $d_{z^2}$ and $d_{x^2-y^2}$ orbitals. With reported Sommerfeld coefficient (9.15 mJ mol$^{-1}$ K$^{-2}$) obtained from specific heat measurement, the estimated DOS($E_F$) $\sim$ 3.88 states/eV/f.u. \cite{schilling-IOP-2008-labanio4} which is about three times larger than that of the DFT results (1.32 states/eV/f.u.). This observation indicates a large quasiparticle mass enhancement due to strong electron correlation in this system. The DFT+DMFT results shown in Fig. \ref{fig1}(c) and (d) follow very similar behaviour as observed in DFT. The feature corresponding to $t_{2g}$ band has almost no influence by electronic correlation and remains at very similar position. On the other hand, $e_{g}$ states are strongly renormalized leading to reduced width of the band and quasiparticle peak moves towards $E_F$ (within $\pm$ 0.2 eV). Mass enhancement factor ($m^*/m_{DFT}$) is weighted sum of contributions arising from all the orbitals weighted by their local Green's function ($\propto$ partial DOS) at $E_F$ \cite{kotliar2DMFT}. Since $t_{2g}$ band does not contribute at $E_F$ and $d_{z^2}$ and $d_{x^2-y^2}$ orbitals (forming $e_g$ band) have almost similar partial DOS at $E_F$; thus we show the real and imaginary part of the average self energy of these orbitals in the inset of Fig. \ref{fig1}(d). Mass enhancement factor is estimated by 1 - $\partial \text{Re}[\Sigma(\omega)] / \partial \omega~|~_{\omega \rightarrow 0}$; which is found to be about 3 in excellent agreement with the experimental results \cite{schilling-IOP-2008-labanio4}. Im[$\Sigma$($i\omega$)] and Re[$\Sigma$($\omega$)] for various values of $U$ and $J_H$ have been shown in Fig. S2 of SM \cite{Supplemental}.

	\begin{figure}[tb]
		\centering
		\includegraphics[width=0.48\textwidth]{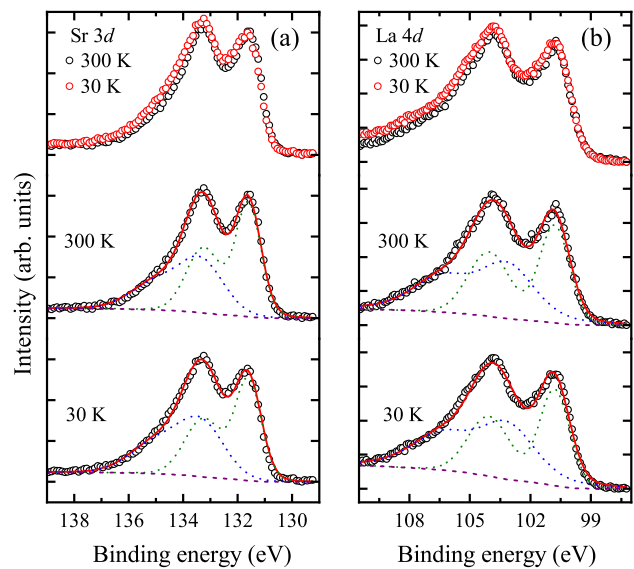}
		\captionsetup{justification=Justified}
		\caption{(color online) (a) Sr 3$d$ and (b) La 4$d$ core level spectra of (La$_{0.5}$Sr$_{0.5}$)$_2$NiO$_4$. Comparison of spectra collected at 300 K and 30 K has been shown in the top panels. Simulation of 300 K and 30 K spectra are shown in middle and bottom panels, respectively. Background, main feature and satellite feature are shown by purple dashed, green dotted and blue dotted lines respectively.}\label{fig2}
\end{figure}

Now we turn our focus on experimental results where Sr 3$d$ and La 4$d$ core level spectra of (La$_{0.5}$Sr$_{0.5}$)$_2$NiO$_4$ have been shown in top panels of Fig. \ref{fig2}(a) and \ref{fig2}(b), respectively. Both the core level spectra were collected using Al $K_\alpha$ radiations at 300 K and 30 K, and the shown spectra have been normalized at lower binding energy feature. The Sr 3$d$ (La 4$d$) spectra exhibit two peaks corresponding to Sr 3$d_{5/2}$ (La 4$d_{5/2}$) and Sr 3$d_{3/2}$ (La 4$d_{3/2}$) positioned at 131.60 eV (100.80 eV) and 133.25 eV (104.05 eV) binding energy, respectively at 300 K. The spin orbit splitting of about 1.65 eV in case of Sr 3$d$ and about 3.25 eV in case of La 4$d$ is apparent from the figure. Intensity ratio of these two features in both, Sr 3$d$ and La 4$d$ spectra, does not follow the expected multiplicity ratio of 3:2 and the appearance of shoulder structure at higher binding energy suggests contribution of more than one spin orbit split doublets \cite{rssingh-APS-2007-Sr3d,broussard-elsevier-1997-La4d,sunding-elsevier-2011-La4d,li-RSC-2019Lacore}. The core hole generated in the photoemission process enhances the local positive charge which can be effectively screened by the valence electrons (in the photoemission final state) leading to well screened feature which appears at lower binding energy, while the ineffective screening of core hole leads to unscreened (satellite) feature at higher binding energy \cite{Hufner}.
	
For quantification, we simulate the Sr 3$d$ spectra at 300 K with two sets of Lorentzian (representing lifetime broadening of the photoholes) convoluted with Gaussian to consider resolution broadening and other solid state effects, where each set represents two spin orbit split features. A Shirley type integral background \cite{shirley-APS-1972} has been shown by a purple dashed line, while green and blue dotted lines represent main and satellite features, respectively. It is to note here that we require larger broadening for the satellite feature than the main feature and spectra could be well fitted as shown in the middle panel of Fig. \ref{fig2}(a). The satellite appears at 1.7 eV higher binding energy having intensity ratio of $\sim$ 70\% with respect to the main feature. Simulation of low temperature spectra, shown in the bottom panel of Fig. \ref{fig2}(a), reveals that the satellite position remains same while the intensity ratio increases to 82\%. A similar analysis has been performed for La 4$d$ spectra, as shown in middle and bottom panels of Fig. \ref{fig2}(b), corresponding to 300 K and 30 K, respectively. Intensity ratio of the satellite feature with respect to the main feature, appearing at 2.35 eV higher binding energy, increases from 90\% to 100\% while going from 300 K to 30 K. In the above analysis procedure, the error in estimating the intensity ratio is smaller than 3\%. The estimated moderate increase in the satellite feature is also evident from the raw spectra, where the higher binding energy feature increases at low temperature. It is to note here that the strength of core hole screening is directly proportional to the number of conduction electrons ($\propto$ DOS($E_F$)). Thus, smaller satellite feature would correspond to larger DOS($E_F$). Moderate increase in the relative intensity of the satellite feature in Sr 3$d$ as well as in La 4$d$ spectra while going from 300 K to 30 K suggests finite reduction of DOS($E_F$) commensurate with transport properties exhibiting MIT \cite{sreedhar-elsevier-1990LSNO,Ytakeda-elsevier-1990LSNO,RJcava-APS-1991LSNO,granados-elsevier-1993-bandgap,ivanova-springer-2002-transport,ishikawa-CSJ-1987-nonstoichiometry,shinomori-JPS-2002MIT,liu-ACS-1991-composition}.
	
	\begin{figure}[tb]
		\centering
		\includegraphics[width=0.48\textwidth]{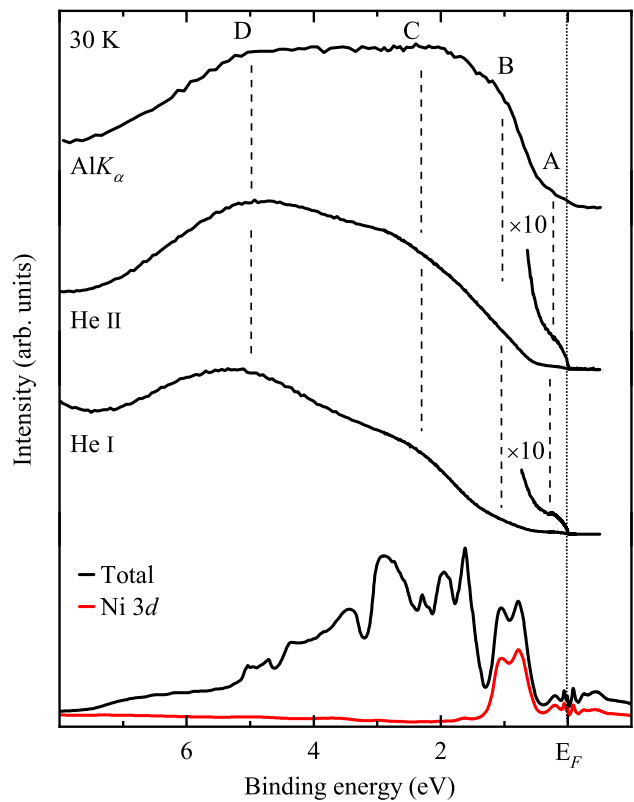}
		\captionsetup{justification=Justified}
		\caption{(color online) Valence band spectra of (La$_{0.5}$Sr$_{0.5}$)$_2$NiO$_4$ collected at 30 K using Al$K_\alpha$, He {\scriptsize II} and He {\scriptsize I} radiations. Distinct features are marked by vertical dashed lines. Intensity of feature A appearing close to $E_F$ is rescaled by $\times$10. Spectral function obtained from DFT+DMFT has also been shown for comparison. All the experimental spectra are shifted vertically for clarity.}\label{fig3}
	\end{figure}
	
	\begin{figure}[tb]
		\centering
		\includegraphics[width=0.44\textwidth]{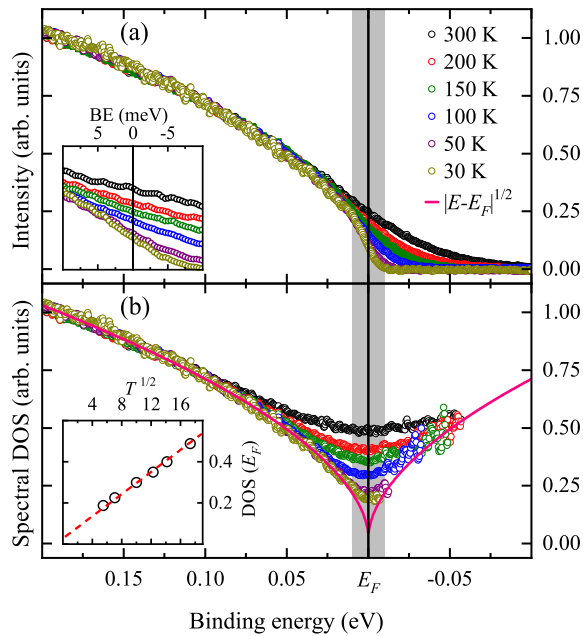}
		\captionsetup{justification=Justified}
		\caption {(color online) (a) High resolution He {\scriptsize I} spectra at different temperatures. Inset shows intensity in the close vicinity of $E_F$. (b) Spectral DOS obtained by division of spectra with resolution broadened Fermi-Dirac distribution function. $|E-E_F|^{1/2}$ dependence is shown by solid line. $T^{1/2}$ dependence of DOS($E_F$) has been shown in the inset.}\label{fig4} 
	\end{figure}
	
	Fig. \ref{fig3} shows the valence band spectra collected at 30 K using Al $K_\alpha$, He~{\scriptsize II} and He~{\scriptsize I} radiations. Similar to the DFT and DFT$+$DMFT results, three distinct sets of features D, C and A+B are observed in all three spectra. Considering larger photoionization cross-section \cite{JJyeh-1985-crosssection} ratio of O 2$p$ states with Ni 3$d$ states in case of He~{\scriptsize II} and He~{\scriptsize I} spectra than that in case of Al $K_\alpha$ spectra, broad features D and C centering around 5 eV and 2.5 eV binding energies, respectively, can be attributed to O 2$p$ states. Similarly, features A \& B appearing below 1.5 eV binding energy can be attributed to Ni 3$d$ states. Comparison of these spectra with the calculated Ni 3$d$ states suggests that the feature B can be ascribed to localized $t_{2g}$ states while feature A represents delocalized $e_g$ states exhibiting small but finite intensity at $E_F$. The observation of finite intensity at $E_{F}$ in the low temperature spectra, where the transport exhibits insulating behaviour \cite{sreedhar-elsevier-1990LSNO,Ytakeda-elsevier-1990LSNO,RJcava-APS-1991LSNO,granados-elsevier-1993-bandgap,ivanova-springer-2002-transport,ishikawa-CSJ-1987-nonstoichiometry,shinomori-JPS-2002MIT,liu-ACS-1991-composition}, suggests the localization of electrons in the vicinity of $E_F$ due to strong disorder present within the system. For clarity, feature A has been shown by $\times$10 scaling in  He~{\scriptsize II} and He~{\scriptsize I} spectra. DFT+DMFT spectral function has also been shown in the figure for a direct comparison with valence band spectra. The peak positions of features B, C and D are captured in both DFT and DFT$+$DMFT calculations, while feature A is represented well by the renormalized $e_g$ band obtained in DFT$+$DMFT results, indicating strongly correlated system.
	
Since the electronic states in the vicinity of $E_F$ are mostly responsible for transport, thermodynamic and other properties of the system, we further show the high resolution He {\scriptsize I} spectra in Fig. \ref{fig4}(a) to understand the temperature evolution of spectral intensity close to $E_{F}$. All spectra, normalized at 200 meV, exhibit finite intensity at $E_F$. While the spectra corresponding to different temperatures remain very similar in larger energy range, the spectral intensity at $E_F$ gradually decreases with decreasing temperature (shown in the inset of Fig. \ref{fig4}(a)). Interestingly, the spectra corresponding to 30 K exhibits a clear Fermi cut-off in contrast to insulating transport at low temperature \cite{sreedhar-elsevier-1990LSNO,Ytakeda-elsevier-1990LSNO,RJcava-APS-1991LSNO,granados-elsevier-1993-bandgap,ivanova-springer-2002-transport,ishikawa-CSJ-1987-nonstoichiometry,shinomori-JPS-2002MIT,liu-ACS-1991-composition}, suggesting Anderson insulating ground state \cite{anderson1958}.
	
In order to clearly visualize temperature induced change close to $E_F$, we show the spectral DOS obtained by dividing photoemission spectra with resolution broadened Fermi-Dirac distribution function at various temperatures. The photoemission intensity can be expressed as $I(E)$ = DOS($E$)$\ast$$F(E, T)$$\ast$$L^e(E)$$\ast$$L^h(E)$$\ast$$G(E)$, where $L^e$ ($L^h$) representing photoelectron (photohole) lifetime broadening which is negligibly small close to $E_F$. $F$ and $G$ represent Fermi-Dirac distribution function and instrumental resolution broadening, respectively. Thus, $I(E)/(F(E, T) \ast G(E))$ provides a good representation of spectral DOS at temperature $T$ as also done in other systems \cite{singh-PRB-2008YIO,*bansal-JPCM-2021LRO,reddy-EPL-2019STIO,Hufner}. Almost identical spectral DOS obtained by symmetrization of the photoemission intensity provides confidence in the analysis (Fig. S3 of SM \cite{Supplemental}). For all the temperatures, the spectral DOS follows $|E-E_F|^{1/2}$ behaviour ($>$ 3$k_BT$), while DOS($E_F$) is found to be monotonically decreasing. Observed values of DOS($E_F$) (0.49, 0.40, 0.35, 0.30, 0.23 and 0.19 at 300 K, 200 K, 150 K, 100 K, 50 K and 30 K, respectively) follow $T^{1/2}$ behaviour as shown in the inset of Fig. \ref{fig4}(b). $|E-E_F|^{1/2}$ dependence of the spectral DOS and the square root temperature dependence of the DOS($E_F$) follows the Altshuler-Aronov theory for the disordered correlated systems \cite{lee-modern-1985disorder,*2021PRBlaSrRhO3films,*kobayashi-PRL-2007AAtheory,*altshuler-elsevier-1979AAtheory}, suggesting strong influence of intrinsic disorder on the electronic structure of (La$_{0.5}$Sr$_{0.5}$)$_2$NiO$_4$. 

 \section*{CONCLUSION}
    Electronic structure of layered nickelate (La$_{0.5}$Sr$_{0.5}$)$_2$NiO$_4$ has been investigated using photoemission spectroscopy and DFT$+$DMFT calculations. Unscreened satellite feature appearing at higher binding energy in Sr 3$d$ and La 4$d$ core level spectra enhances with lowering of temperature, suggesting reduction of DOS($E_F$) commensurate with transport measurements. Valence band spectra could be well captured within DFT$+$DMFT calculations indicating strong electronic correlation. These calculations also suggest strong renormalization of partially filled $e_g$ band with $m^*/m_{DFT} \sim$~3, which are in excellent agreement with specific heat measurements. Room temperature high resolution valence band spectra reveals small intensity at $E_F$, which evolves to exhibit a sharp Fermi cut-off at 30 K. This indicates metallic character of (La$_{0.5}$Sr$_{0.5}$)$_2$NiO$_4$ in contrast to insulating transport at low temperatures suggesting Anderson insulating ground state. Spectral DOS follows $|E-E_F|^{1/2}$ dependence, suggesting disorder induced localization of electrons. The square root temperature dependence of DOS($E_F$) follows the Altshuler-Aronov theory for the disordered correlated systems.

 \section*{ACKNOWLEDGMENTS}
    We acknowledge the support of Central Instrumentation Facility and HPC Facility at IISER Bhopal. The support from DST-FIST (Project No. SR/FST/PSI-195/2014(C)) is also thankfully acknowledged.
	
	\providecommand{\noopsort}[1]{}\providecommand{\singleletter}[1]{#1}%

\end{document}


\preprint{preprint}

\title {Supplementary Material for ``Role of electron correlation and disorder on the electronic structure of layered nickelate (La$_{0.5}$Sr$_{0.5}$)$_2$NiO$_4$"}

\author{Sakshi Bansal}
\affiliation{Department of Physics, Indian Institute of Science Education and Research Bhopal, Bhopal Bypass Road, Bhauri, Bhopal 462066, India}%

\author{R. K. Maurya}
\affiliation{Department of Physics, Indian Institute of Science Education and Research Bhopal, Bhopal Bypass Road, Bhauri, Bhopal 462066, India}%

\author{Asif Ali}
\affiliation{Department of Physics, Indian Institute of Science Education and Research Bhopal, Bhopal Bypass Road, Bhauri, Bhopal 462066, India}%

\author{B. H. Reddy}
\altaffiliation[Present address: ]{Department of Physics, Government College (A), Rajahmundry 533105, India}
\affiliation{Department of Physics, Indian Institute of Science Education and Research Bhopal, Bhopal Bypass Road, Bhauri, Bhopal 462066, India}%

\author{Ravi Shankar Singh}
\email{rssingh@iiserb.ac.in}
\affiliation{Department of Physics, Indian Institute of Science Education and Research Bhopal, Bhopal Bypass Road, Bhauri, Bhopal 462066, India}%

\date{\today}

\maketitle
\begin{figure}[h]
	\centering
	\includegraphics[width=0.75\textwidth]{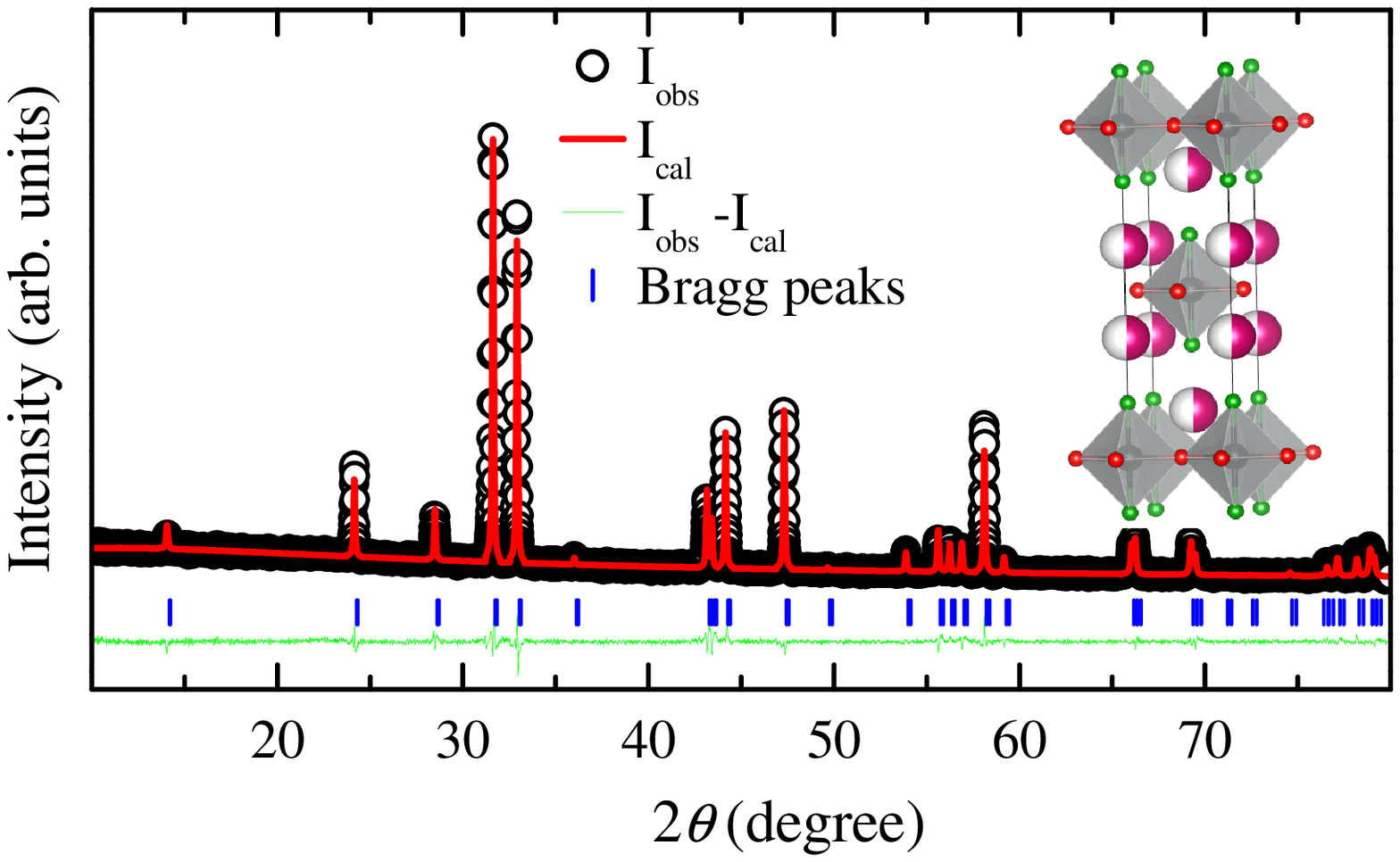}
	\vspace{-72ex}
	\captionsetup{justification=Justified}
	\caption *{\label{fig5} Fig. S1: (color online) $X$-ray diffraction pattern (open circles) and Rietveld refinement results of (La$_{0.5}$Sr$_{0.5}$)$_2$NiO$_4$, where red line, blue ticks and green lines show the calculated pattern, Bragg peak positions and the difference spectra, respectively. Inset shows the crystal structure where pink/white, black, green and red spheres represent La/Sr, Ni, O1 (apical) and O2 (planner) atoms, respectively.}
\end{figure}

\begin{figure}[h]
	\centering
	\includegraphics[width=0.75\textwidth]{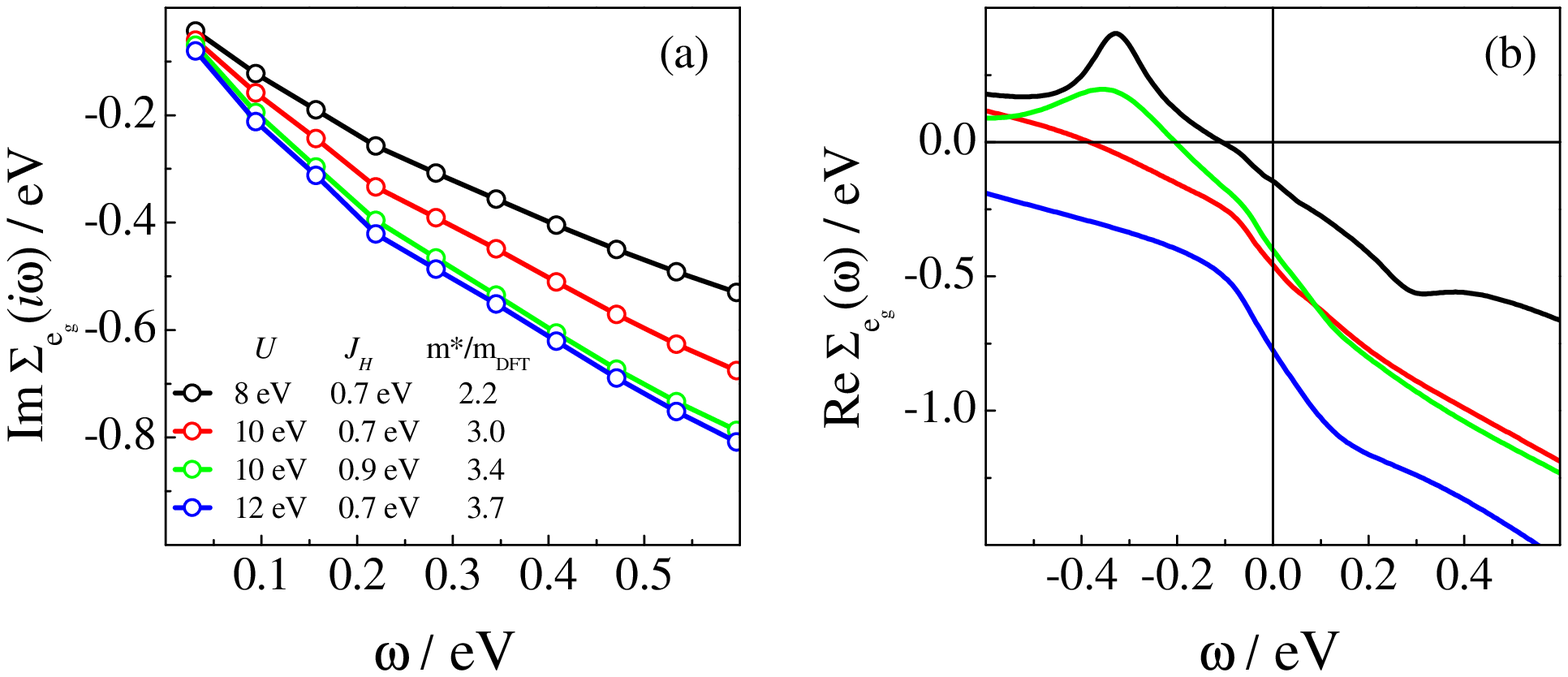}
	\vspace{-82ex}
	\captionsetup{justification=Justified}
	\caption *{\label{fig5} Fig. S2: (color online) (a) Imaginary part of the average self energy of $e_g$ bands on imaginary frequency axis, (b) real part of the average self energy of $e_g$ bands on real frequency axis for various values of $U$ and $J_H$. $m^*/m_{\text{DFT}}$ increases with increasing $U$ as well as $J_H$.}
\end{figure}
\vspace{-4ex}
\begin{figure}[h]
	\centering
	\includegraphics[width=0.75\textwidth]{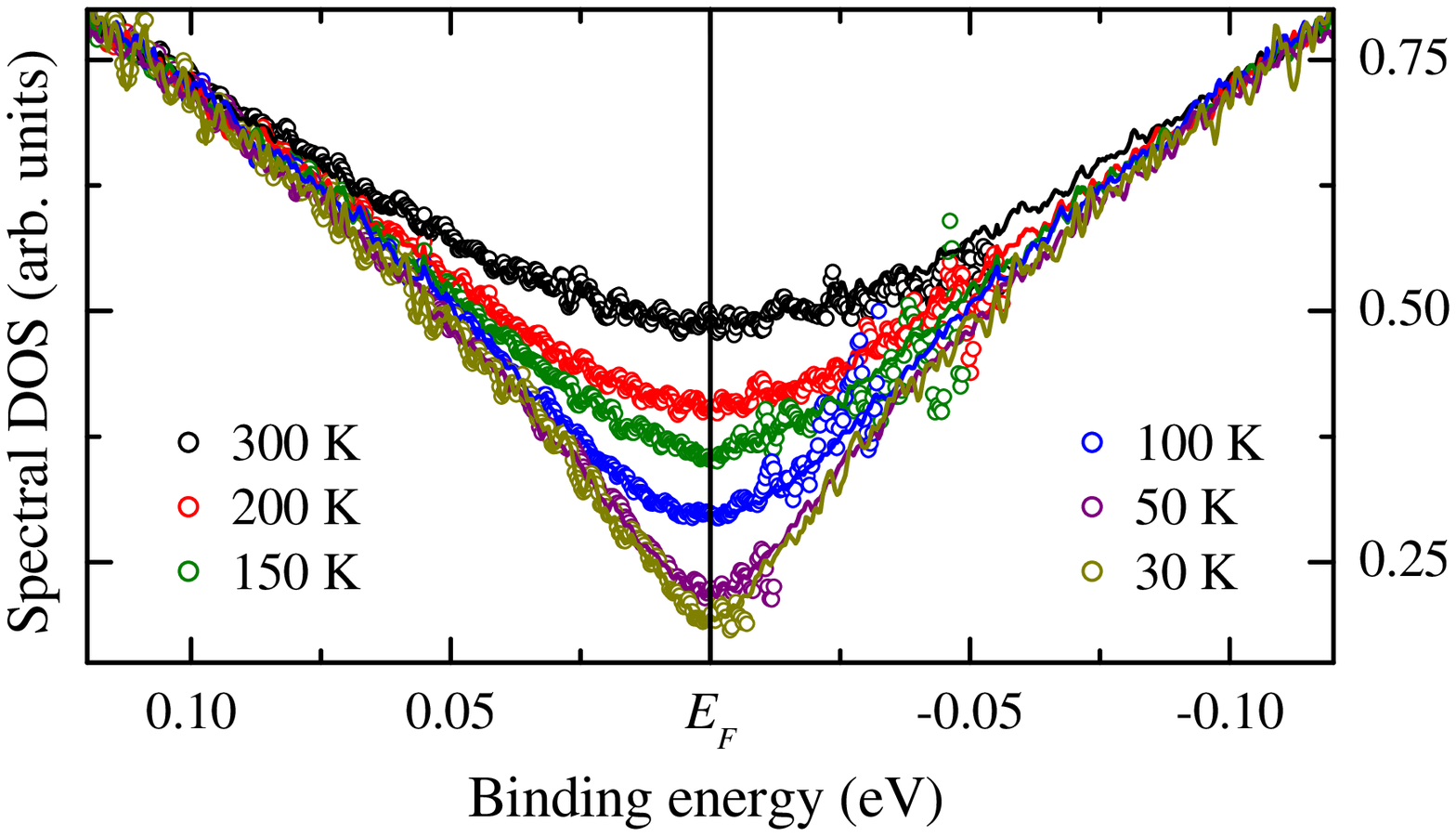}
	\vspace{-70ex}
	\captionsetup{justification=Justified}
	\caption *{\label{fig5} Fig. S3: (color online) Comparison of spectral DOS obtained using two methods for various temperatures. Open circles show the spectral DOS obtained by dividing the spectra with resolution broadened Fermi-Dirac distribution function (same as main text), and line represent the same obtained by symmetrization of the spectra. In case of symmetrization, $I(E) +I(-E)$ provides the description of spectral DOS, assuming DOS is not abruptly changing around $E_F$, and Fermi-Dirac distribution function (also resolution broadened) follows the relation $F(E) +F(-E)$~=~1.}
\end{figure}